
\def\gff{\hbox{l}\!\!I}
\def\pd{\partial}
\def\dt{\hbox{det}}
\magnification=1200
\parskip 3 pt plus 1pt minus 1 pt
\rightline{DTP-91/55}
\rightline{ITEP-M-6/91}
\rightline{October, 1991}
\vskip 2 true cm
\centerline{UNIVERSAL FIELD EQUATIONS WITH COVARIANT SOLUTIONS}
\vskip 2.5 true cm
\centerline{D.B. FAIRLIE,\ J. GOVAERTS,\ A.  MOROZOV%
\footnote{$~{\dag}$}
{\it  Also Institute for Theoretical Physics, University of Helsinki, Finland.}
\footnote{$~{\ddag}$}{\it On leave of absence from I.T.E.P. 117259, Moscow.}}

\vskip 0.5 true cm
\centerline{\it{Department of Mathematical Sciences}}
\centerline{\it{University of Durham, Durham DH1 3LE, England}}
\vskip 2 true cm
\centerline{Abstract}
 \vskip 1 true cm

Metric independent $\sigma$ models are constructed. These are
 field theories which generalise the membrane idea to situations
where the target space has fewer dimensions than the base manifold.
Instead of reparametrisation invariance of the
independent variables, one has invariance of solutions of the
field equations under arbitrary functional redefinitions of the field
quantities. Among the many interesting properties of these new models is the
existence of a hierarchical structure which is illustrated by the following
result.

Given an arbitrary Lagrangian, dependent only upon first
derivatives of the field, and homogeneous of weight one, an iterative
procedure for calculating a sequence of equations of motion is discovered,
which ends with a universal, possibly integrable equation, which is
 independent of the starting Lagrangian. A generalisation  to more than
one field is  given.

\vfill\eject
\vskip 10pt
\centerline{\bf 1. Introduction}
\vskip 10pt
   This paper is devoted to the study of metric independent equations
 which are invariant under general field redefinitions.
In various speculations about quantum gravity it seems desirable to change
classical co-ordinates in space time for quantum operators. A step in this
direction is provided by $\sigma$-models, i.e. $d$-dimensional field theories
with $\cal D$ fields $\phi~a (x_i),\ (a=1,\dots {\cal D},\ i=1,\dots,d)$
interpreted as a map from ${I\!\!R }~d$ into some $\cal D$-dimensional manifold
${\cal M~D}$. However the $\sigma$ model action depends explicitly on the
metric on ${\cal M~D}$ and in order to restore $\cal D$-dimensional covariance
with respect to any general co-ordinate transformation
$$
\phi~a\rightarrow {\cal F}~a(\phi_1\dots\phi_{\cal D})\equiv {\cal F}\{\phi\},
\eqno(1.1)$$
one should treat the $\sigma$-model metric $G_{ab}\{\phi\}$ as a new quantum
field. This kind of approach involving the ``second quantisation'' of $\sigma$
models, is a piece of the standard string program. It seems however natural to
wonder whether one can obtain general covariance in a more direct manner, by
the use of equations of motion invariant under field redefinitions, whose
solutions as functions of the base manifold of dimension $d$ may thus  be
interpreted as models of space-time co-ordinates. In a sense, the equations
we shall construct possibly represent a halfway house between the ordinary,
and topological field theories$~{[1]}$\rlap, with field redefinitions as a
gauge
invariance. One possible approach is based upon the notion of a
dualisation between reparametrisation invariance and invariance under
field redefinition. The former holds for
string and membrane theory and its higher dimensional generalisations, i.e.
$d$ dimensional manifolds embedded in a target space
of $\cal D$ dimensions, with ${\cal D} > d$. The latter obtains in the
natural extension of these theories to the case where ${\cal D} < d$.

As is well known, when ${\cal D} > d$   the Lagrangian density
is taken as simply the natural generalisation of the Nambu-Goto
action$~{[2,3]}$ for strings, namely the volume density of the embedded
 manifold.
If $\phi~a(x_j)$ where $a = 1,2,\dots,{\cal D}$ and $j = 1,2,\dots,d$,
then this Lagrangian density is simply
$$ {\cal L}_{NG}  =
\sqrt{\hbox{det}\bigl(\sum_{a=1}~{\cal D}
{\pd\phi~a\over\pd x_j}{\pd\phi~a\over\pd x_k}\bigr)}.\eqno(1.2)
$$
The appealing feature of this Lagrangian is that the resulting action is
invariant under reparametrisations of the co-ordinates of the base manifold,
i.e. is invariant under diffeomorphisms. In the case of
$d=2$ this invariance is enough to guarantee that the theory is
integrable.

 In this article we explore among other ideas, an  analogous generalisation
to the case where ${\cal D} < d$. This means that the target space is of lower
dimension than the space of independent variables. Thus this situation is more
like that of a field theory. We take the metric of the base manifold to be
flat, and either Euclidean or Lorentzian.
Consider the Lagrangian density
$$ {\cal L} =
\sqrt{\hbox{det}\bigl(\sum_{j=1}~d
{\pd\phi~a\over\pd x_j}{\pd\phi~b\over\pd x_j}\bigr)}.\eqno(1.3)
$$
This Lagrangian is invariant under orthogonal (Lorentzian)
transformations of ${I\!\!R}~d$ and, as we shall show, leads to equations of
motion invariant under all field redefinitions of the dependent fields
$\phi~a$. In other words, if $\phi~a$ is a solution to the equations of
motion, so is $F~a(\phi~b)$, where $F~a(\phi~b)$ are an arbitrary
functions! This remarkable property is shared by a surprisingly large class of
Lagrangians which we describe later. Notice that the Lagrangian itself is
not invariant, only covariant under such transformations. The relation between
homogeneity of the Lagrangian, and covariance of the field equations is
elucidated in section 3.  The connection between
(1.2) and (1.3) becomes more transparent with the introduction of the
following notation.
$$ \hbox{Let matrix} \quad N=\quad N_{a j}=  {\pd \phi~a\over\pd x_j}.$$
Then first case the  Lagrangian (1.2) is simply
$${\cal L}_{NG} =    \sqrt{(\hbox{det}\tilde N N)}, $$
where $\tilde N$ denotes the matrix transpose.
Here the inner product of $\tilde N$ with $N$ is non-singular,
whereas for (1.3) the Lagrangian is
$${\cal L}=\sqrt{(\hbox{det}N\tilde N)}$$
since   it is the outer product which is non singular in this case.

The case where ${\cal D} =1,\  d=2$ is that of
the so called Bateman equation$~{[3-5]}$\rlap,
$$\phi_{x_1x_1}\phi_{x_2}~2+\phi_{x_2x_2}\phi_{x_1}~2
-2\phi_{x_1x_2}\phi_{x_1}\phi_{x_2} =0.\eqno(1.4)$$
Its general solution has
been known for some time,  and is given by the solution
for $\phi$ of the implicit equation
$$ x_1f~1(\phi) + x_2f~2(\phi) =c \quad\hbox{constant.}\eqno(1.4)$$
where $f~1,\  f~2$ are any two arbitary functions of $\phi$, and $c$
may generally be taken as  one  or zero.
In fact a class of solutions to the problem with
Lagrangian (1.3) may be obtained as the implicit solution of a vector valued
generalization of (1.4), namely

$$ \sum_{j=1}~d    x_jf~{j,a}(\phi~{b})=   c~{a}\eqno(1.5)$$

We shall begin in section 2 by reviewing known facts about the Bateman
equation, and exhibit different ways in which it can be represented.
 It turns out that the equation (1.4) can be derived from infinitely many
Lagrangians which are not related by divergences, and are thus
inequivalent. It is this feature which accounts for the integrability of the
Bateman equation.

After the discussion of homogeneity and covariance in section 3, the main
result
of this paper is contained in the theorem of section 4. We prove that if
${\cal L}$ is a Lagrangian density dependent only upon first derivatives of the
field, which is homogeneous of weight one, and ${\cal E}$ is the Euler
derivative, then there is a sequence of  equations of motion
$({\cal EL})~k =0\ (k=1,\dots,d)$ such that $({\cal EL})~d\equiv0$ and for
$k=d-1$
is independent of the starting Lagrangian, and is the universal equation of the
title. The multitude of Lagrangians at level $k=d-2$  which produce this
equation may be regarded as equivalent to conservation laws, hence the
prospect of complete integrability of the universal equation.
All members of the hierarchy in fact share the common solution (1.5), and the
procedure for calculating the iterative sequence is related to de Rahm
cohomology.

In section 5 we generalise these considerations to ${\cal D}>1$. If
${\cal D}=d-1$, then we obtain ${\cal D}>1$ universal equations for the fields
$\phi~a,\ a=1,\dots,{\cal D}$. We have not as yet found an analogous
hierarchical procedure for the general case, but  propose an overdetermined
set of ${d-1  \choose{\cal D}-1}$ equations as the universal set. This set is
non-empty, as all members share the common solution (1.5). The idea that
physically relevant quantities might be specified by an overdetermined set of
equations is not altogether new, as the classical string solutions  themselves
 are determined by the class  of solutions of the free wave equation subject to
two quadratic constraints. The paper is then rounded off with some remarks in
conclusion.
\vskip 10pt

\leftline{\bf 2. The Bateman equation}
 \vskip 10pt
This equation first appeared in an article by Bateman$~{[4,5]}$
in a hydrodynamical context. It is a specialisation to infinite
light velocity of the well known Born-Infeld equation and in $x,t$ co-ordinates
takes the form
$$\phi_{xx}\phi_t~2+\phi_{tt}\phi_x~2-2\phi_{tx}\phi_x\phi_t =0.\eqno(2.1)$$
This equation can be viewed in various ways;
\item{A)} As a determinantal equation:
$$ \hbox{det}\pmatrix{0&\phi_x&\phi_t\cr
                      \phi_x&\phi_{xx}&\phi_{xt}\cr
                      \phi_t&\phi_{tx}&\phi_{tt}\cr}=0 \eqno(2.2) $$
In this form it is easy to see that the equation is invariant under not only
Euclidean (Lorentz) transformations on the co-ordinates $x,t$ but also under
the full general linear group $GL(2,R)$. In addition the remarkable property
of invariance  under replacement of any solution $\phi(x,t)$ of (2.2)
by any arbitrary twice differentiable function $F(\phi)$ is also manifest.
\item{B)} Equation (2.1) is also derivable from the Laplace (wave) equation in
3 dimensions , subject to the non-linear constraint that the gradient of $\phi$
is a null vector:
$$\eqalignno
{\phi_{xx}+\phi_{yy}\pm\phi_{tt} &=0.&(2.3a)\cr
 \phi_x~2+\phi_y~2\pm\phi_t~2&=0.&(2.3b)\cr}$$
after    elimination of $\phi_y,\  \phi_{yy}$.
Again, the property of redefinition of $\phi$ is easy to see. Notice that in
this formulation, the left hand side of (2.3b) is the Lagrangian for (2.3a).
However, the equation of motion has to be evaluated on the subspace where the
Lagrangian itself vanishes! This is a hint of a possible connection with
topologoical field theories
\item{C)} Another formulation in which the property of field redefinition is
evident comes from the observation that equation (2.1) results from the
substitution of $u(x,t)={\phi_t\over\phi_x}$ in the first order differential
equation$~{[5]}$
$$ {\pd u\over\pd t} = u{\pd u\over\pd x}.\eqno(2.4)$$
This equation, which is the K dV equation without the term in $u_{xxx}$,
admits an  interpretation in 2 dimensional fluid mechanics as
the vanishing of the Lagrangian derivative of the velocity field.
This formulation is especially convenient to reveal integrable structure.
It is easy to see that (2.4) possesses an infinite number of conservation
laws since trivially (2.4) implies
$$  {\pd\over\pd t}(u~n)={\pd\over\pd x}({n\over n+1}u~{n+1}).\eqno(2.5)$$

In fact the general solution, in the form of an implicit
equation for $u(x,t)$ is well known:
$$ u= G(x+ut),\quad\hbox{where}\   G(x)\ \hbox{is arbitrary}.\eqno(2.6)$$

The generic solution of the Bateman equation can be deduced from this, or else
verified ab initio. Take two arbitrary functions $F_1,\ F_2$ which
depend only upon $\phi(x,t)$ and subject them to the single constraint
$$ xF_1(\phi) +tF_2(\phi)=c.\eqno(2.7)$$
Here $c$ may be chosen as zero, or 1.
Then regarding this as an implicit equation for $\phi$, it is easy to
verify that (2.1), taken in the form (2.2) is solved.  The Cauchy problem
for this equation is not always well posed. However, in the generic case
given initial and final
boundary values $\phi(x,t)|_{t=0} =g_0(x)$ and $\phi(x,t)|_{t=1}=g_1(x)$
and taking $c=1$,
then the two functions $F_1$ and $F_2$ are defined uniquely from the two
equations
$$xF_1(g_0(x))=1;\quad\quad F_2(g_1(x)) =1- F_1(g_1(x)),\eqno(2.8)$$
provided neither $g_0(x)$ nor $g_1(x)$ are constant.

The final topic in this section concerns the existence of a Lagrangian for
the Bateman equation. Because of the invariance of the equation under
$GL(2,R)$ any Lagrangian of the form
$$\sqrt{a\phi_x~2+2b\phi_x\phi_t+c\phi_t~2},\quad b~2-ac \neq 0.\eqno(2.9)$$
will do! Another curiosity which stems from this observation is that the
Bateman equation remains form invariant under a transformation to
holomorphic (light-cone) co-ordinates. This remark explains how an evidently
Lorentz invariant equation can arise from one which is apparently not
invariant i.e. (2.4). This equation in fact is invariant when $x$ and $t$
are treated as light-cone co-ordinates, and $u$ is appropriately scaled.
$x'\rightarrow \lambda x;\ t'\rightarrow {1\over\lambda}t;\
u'\rightarrow \lambda~2u$
The Lagrangian (2.8) scales under field redefinition. Our next task is to
investigate how these properties can be generalised to larger numbers of
fields, and more co-ordinates. Two special cases of this Lagrangian are
noteworthy;

$$\sqrt{\phi_x~2+\phi_t~2},\quad\hbox{and}\quad\sqrt{\phi_x\phi_t}.\eqno(2.10)$$
The first of these leads to a higher dimensional generalisation
of the type discussed in the Introduction (1.2). In fact the class of
admissible Lagrangians is  much larger than (2.8), as we shall explain in
section 4,
being all functions ${\cal L}(\phi_x,\phi_y)$ such that not all second
partial derivatives vanish (non vanishing Hessian) and such that  they are
homogeneous of weight 1, i.e.
$${\cal L}(\lambda\phi_x,\lambda \phi_y)=\lambda{\cal L}(\phi_x,\phi_y).
\eqno(2.11)$$
\vskip 10pt
\leftline{\bf 3. Homogeneity and covariance.}
 \vskip 10pt
Consider first of all a Lagrangian which depends upon a scalar field $\phi$
only through its first derivatives: ${\cal L}(\phi_j)$. The Euler-Lagrange
equation of motion is simply
$$\sum_i\pd_i{\pd{\cal L}(\phi_j)\over\pd\phi_i}=0.    \eqno(3.1)$$
Suppose also that ${\cal L}(\phi_j)$ is homogeneous of weight $\alpha$, i.e.
$${\cal L}(\lambda \phi_j)=\lambda\sp\alpha{\cal L}(\phi_j)\eqno(3.2)$$
This implies that ${\cal L}$ satisfies the Euler equation
$$\sum_i\phi_i{\pd{\cal L}(\phi_j)\over\pd\phi_i}=\alpha{\cal L}(\phi_j)
\eqno(3.3)$$
and that ${\pd{\cal L}(\phi_j)\over\pd\phi_i}$ is a homogeneous function of
weight $\alpha-1$. Then we have the following result:
\smallskip
\leftline{\it Theorem.}
\smallskip
 Suppose ${\cal L}(\phi_j)$ is  homogeneous of weight $\alpha$, and that
$\phi(x_j)$ is a solution of the equation of motion (3.1). Then the condition
that any function ${\Phi}= F(\phi)$ is also a solution to (3.1), i.e.
the equation of motion is covariant under field redefinition is simply
$\alpha(\alpha-1)=0$.
\smallskip
\leftline{\it Proof.}
\vskip 10pt

Consider ${\mit\Phi}= F(\phi),\ {\mit\Phi}_i= F\sp\prime(\phi)\phi_i$. Then;
$$\eqalign{
\sum_i\pd_i{\pd{\cal L}({\mit\Phi_j})\over\pd\phi_i}=&
\sum_i\pd_i{\pd{\cal L}(F\sp\prime \phi_j)\over\pd\phi_i}\cr
=&\sum_i\pd_i\bigl[(F\sp\prime)\sp{\alpha-1} {\pd{\cal
L}(\phi_j)\over\pd\phi_i}
\bigr]\cr
=&(F\sp\prime)\sp{\alpha-1}\sum_i\pd_i{\pd{\cal L}(\phi_j)\over\pd\phi_i}
\quad(=0)\cr
 &+(\alpha-1)(F\sp\prime)\sp{\alpha-2}F\sp{\prime\prime}
  \sum_i\phi_i{\pd{\cal L}(\phi_j)\over\pd\phi_i}\cr
 =&\alpha(\alpha-1)(F\sp\prime)\sp{\alpha-2}F\sp{\prime\prime}
{\cal L}(\phi_j).\cr}\eqno(3.4)$$
Hence if $\alpha=0$  or $\alpha=1$ the equation of motion is covariant.

 There is a class of solutions to (3.1) which is a direct generalisation of the
Bateman solution (2.6) in the case where $\alpha=1$. Suppose $F_i$ are
arbitrary functions of $\phi$ subject to the single constraint
$$\sum_{i=1}\sp dx_iF_i(\phi)=c \quad\hbox{(constant)}.\eqno(3.5)$$
Then
$$\eqalign{
\phi_i=&{-F_i\over\sum_i x_iF_i\sp\prime}\quad
{\buildrel\rm def\over=}\quad {-F_i\over S},\cr
\phi_{ij}=&{F_i\sp\prime F_j
+F_iF_j\sp\prime\over(\sum_ix_iF_i\sp\prime)\sp2}
+{F_iF_j(\sum x_kF_k\sp{\prime\prime})\over(\sum_i x_iF_i\sp\prime)\sp3}\cr
=&{F_i\sp\prime F_j +F_iF_j\sp\prime\over S\sp2}+{F_iF_jS\sp\prime\over S\sp3}
\cr} \eqno(3.6)$$
These relations imply that $\phi_{ij}=f_i\phi_j+f_j\phi_i$ for some functions
$f_i$, a representation which is useful in verifying this solution of the
the hierarchy of equations in section 4.
Inserting the relations (3.6) into the equation of motion  (3.1) we obtain
$$\eqalign{
 \sum_i\pd_i{\pd{\cal L}\over\pd\phi_i}({-F_j\over S})=&
 \sum_i\pd_i\bigl[({-1\over S})\sp{\alpha-1}
  {\pd{\cal L}\over\pd\phi_i}(F_j)\bigr]\cr
=&(\alpha-1)({-1\over S})\sp{\alpha}\sum_i[F_i\sp\prime -{S\sp\prime
\over S}F_i]{\pd{\cal L}\over\pd\phi_i}(F_j)+
({-1\over S})\sp{\alpha}\sum_{i,k}F_iF_k\sp\prime{\pd\sp2{\cal L}
\over\pd\phi_i\pd\phi_k}(F_j).\cr}\eqno(3.7)$$
Now when $\alpha =1$ the first term vanishes, and since
$$\sum_iF_i {\pd{\cal L}\over\pd\phi_i}(F_j)=\alpha{\cal L}(F_j),\eqno(3.8)$$
$$\sum_iF_i\sp\prime{\pd{\cal L}\over\pd\phi_i}(F_j)
+\sum_{i,k}F_iF_k\sp\prime{\pd\sp2{\cal L}\over\pd\phi_i\pd\phi_k}(F_j)
=\alpha\sum_iF_i\sp\prime{\pd  {\cal L}\over\pd\phi_i}(F_j).\eqno(3.9)$$
Rearranging:
$$ \sum_{i,k}F_iF_k\sp\prime{\pd\sp2{\cal L}\over\pd\phi_i\pd\phi_k}(F_j)
=( \alpha-1) \sum_iF_i\sp\prime{\pd{\cal L}\over\pd\phi_i}(F_j).\eqno(3.10)  $$
Thus the second term in the final expression also vanishes when $\alpha=1$,
and the claimed solution is verified.

 These results may be readily extended to the case where many fields
$\phi\sp a$ are present. Suppose the Lagrangian depends only upon first
derivatives of the fields. Then the equations of motion are
$$\sum_i\pd_i{\pd{\cal L}(\phi\sp b_j)\over\pd\phi\sp a_i}=0.    \eqno(3.11)$$
Assume that under  field redefinitions the Lagrangian transforms as
$$\eqalign{
{\cal L}(\phi\sp b_iA\sp{a}_b)=&(\hbox{det}A)\sp\alpha{\cal L}(\phi\sp{a}_i),
\ \hbox{which implies}\cr
\sum_i\phi\sp a_i{\pd{\cal L}\over\pd\phi\sp b_i}(\phi\sp a_j)=&
           \alpha\delta\sp a_b{\cal L}(\phi\sp a_j),\cr
{\pd{\cal L}\over\pd\phi\sp a_i}(\phi\sp b_iA\sp a_b)=&(\hbox{det}A)\sp\alpha
(A\sp{-1})\sp b_a{\pd{\cal L}\over\pd\phi\sp b_i}(\phi\sp
a_i).\cr}\eqno(3.12)$$

If $\Phi\sp a=F\sp a(\phi\sp b)$ then
$$\Phi_i\sp a=\phi_i\sp bA\sp a_b(\phi\sp a),\quad\hbox{with}\quad
 A\sp a_b(\phi\sp a)=\pd_b F\sp a(\phi\sp a),\eqno(3.13)$$
so that
$$\eqalign{\sum_i\pd_i{\pd{\cal L}\over\pd\phi_i\sp a}(\phi_j\sp bA_b\sp a)
&=\pd_i\bigl[(\dt A)\sp\alpha(A\sp{-1})_a\sp b{\pd{\cal L}\over\pd\phi_i\sp b}
(\phi_j\sp a)\bigr]\cr
&=(\dt A)\sp\alpha\bigl[\alpha(A\sp{-1})_d\sp c(A\sp{-1})_a\sp b
-(A\sp{-1})_a\sp c(A\sp{-1})_d\sp b\big]\alpha\pd_b\pd_cF\sp d
{\cal L}(\phi\sp a_j)\cr
&=          \alpha(\alpha-1)( \dt A)\sp\alpha(A\sp{-1})_a\sp b(A\sp{-1})_d\sp c
      \pd_b\pd_cF\sp d {\cal L}(\phi\sp a_j).\cr}\eqno(3.14)$$
Hence; if $\alpha(\alpha-1)=0$ the equations are covariant under
   $\phi\sp a\mapsto{\mit \Phi}\sp a=F\sp a(\phi\sp a).$

We shall conclude this section with a brief mention of the case ${\cal D}=1$
where $\cal L$ depends upon second derivatives $\phi_{ij}$ linearly, in
addition to an arbitary homogeneous  dependence  upon $\phi_j$, i.e.
$$\eqalign{
{\cal L}(\phi_i,\phi_{ij})=& \sum_{i,j}{\cal F}_{ij}(\phi_i)\phi_{ij},\cr
\hbox{with}\quad{\cal F}_{ij} (\lambda\phi_k)=&\lambda~\alpha{\cal F}_{ij}
(\phi_k),\quad{\cal F}_{ji}(\phi_k) = {\cal F}_{ij}(\phi_k).\cr}\eqno(3.15)$$
Then, an analysis of the equation of motion shows that it is covariant under
   $\phi\mapsto{\mit\Phi}=F(\phi),\forall F(\phi)$,
provided that the functions ${\cal F}_{ij}(\phi_i)$ obey the equations
$$\sum_{k,l}\phi_k\phi_l\bigl[{\pd~2{\cal F}_{kl}\over\pd\phi_i\pd\phi_j}
-{\pd~2{\cal F}_{kj}\over\pd\phi_i\pd\phi_l}
-{\pd~2{\cal F}_{ki}\over\pd\phi_j\pd\phi_l}\bigr]=\alpha(\alpha-1){\cal
F}_{ij}
.\eqno(3.16)$$
  Making different ans\"atze for the functions ${\cal F}_{ij}$, it is
straightforward to find explicit solutions to this equation. This procedure can
evidently be generalised to encompass Lagrangians with higher derivatives.
\vskip 10pt
\leftline{\bf 4. Iterative Lagrangians and the Universal Equation.}
\vskip 10pt
In this section we come to the heart of the matter, and show that there is
a special covariant equation in each dimension, which arises as a result of
an iterative construction of a hierarchy of equations, starting from an
arbitrary generic Lagrangian. Among  the remarkable features of this
construction are the consequences that the equations involve nothing further
than second derivatives of the field $\phi$ and although the starting
Lagrangian need have no symmetry, the final equation is $GL(d)$ invariant.

 Suppose ${\cal L}_0(\phi_j),\ j=1,\dots d$ depends only upon first derivatives
of the field $\phi$, and is  homogeneous of weight one, and the matrix
$M_{ij}=  {\pd~2{\cal L}_0(\phi_k)\over\pd\phi_i\pd\phi_j} $ is of maximal rank
$d-1$, which is the generic case.
Denote by ${\cal E}$ the Euler differential operator
$${\cal E}=-{\pd\over\pd\phi}
 +\pd_i {\pd\over\pd\phi_i}-\pd_i\pd_j{\pd\over\pd\phi_{ij}}\dots
\eqno(4.1)$$
(In principle the expansion continues indefinitely  but it is sufficient for
our purposes to terminate at the stage of second derivatives  $\phi_{ij}$).
Note the following property of ${\cal E}$:
\vskip 10pt
 \leftline{\it Lemma 1.}
\smallskip
$ {\cal E}~2 {\cal K}(\phi_i,\phi_{ij},\dots)= 0,\ \forall {\cal K}$.
This follows
simply from the observation that ${\cal E}{\cal K}$ is then a divergence,
thus when treated as a Lagrangian leads to a null equation of
motion  \rlap,$~{[6]}$ or else by direct calculation. Thus ${\cal E}$ is in
the nature of a BRST operator.

Now consider the sequence of equations of motion;
$$\eqalign{ {\cal E\ L}_0 &=0\cr
            {\cal E\ L}_0{\cal E\ L}_0 &=0\cr
{\cal E\ L}_0{\cal E\ L}_0{\cal E\ L}_0 &=0\quad\hbox{etc.}\cr}\eqno(4.2)$$
Then this sequence terminates after $d$ iterations when the left hand side
vanishes identically. At the penultimate step the resulting equation of motion
is universal; i.e. is independent of the details of ${\cal L}_0$, and is in
fact
proportional to the equation
$$\det\pmatrix{0&\phi_1&\phi_2&\ldots&\phi_d\cr
               \phi_1&\phi_{11}&\phi_{12}&\ldots&\phi_{1d}\cr
               \phi_2&\phi_{12}&\phi_{22}&\ldots&\phi_{2d}\cr
                    .&\       .&\       .&\ddots&\       .\cr
               \phi_d&\phi_{1d}&\phi_{2d}&\ldots&\phi_{dd}\cr}=0,\eqno(4.3)$$
or, more succinctly
$ \sum_{i,j}\phi_i |\phi_{kl}|~{-1}_{ij}\phi_j=0$.
 These results may be established by two methods ; an iterative proof, and
alternatively by the construction of a generating function.
\vskip 10pt
\leftline{\it Iterative proof.}
\vskip 10pt
Define an iterative sequence
$$\eqalign{
{\cal L}_{n+1}=&-\lim_{\epsilon \rightarrow 0}{({\cal L}_n(\phi+\epsilon V)-
{\cal L}_n(\phi))\over\epsilon}+
{\pd\over\pd x_k}({\cal L}_n{\pd V\over\pd\phi_k}) \cr
=&V{\cal E}{\cal L}_n +\  \hbox{total derivative},\cr}\eqno(4.4)$$
where ${\cal L}_n$ and $V$ arbitrary functions of $\phi$ and its derivatives.
The equivalence between those two expressions is a consequence of the fact
that only terms linear in $\epsilon$ and hence $V$ in
$({\cal L}_n(\phi+\epsilon V)-{\cal L}_n(\phi))$ ever contribute. Note that in
the case where ${\cal L}_n$ and $V$ are both independent of $\phi$, but not of
its derivatives, the total derivative in (4.4) does not contribute to
${\cal EL}_{n+1}$ in consequence of Lemma 1. The sequence (4.4) possesses
two main properties. If $V$ depends only upon $\phi_i$, and ${\cal L}_n$ only
upon $\phi_i$ and $\phi_{ij}$, then the terms
in $\phi_{ijk}$ arising from the first term in (4.4), namely
$({\pd{\cal L}_n\over\pd\phi_{ij}})({\pd~2 V\over\pd x_i\pd x_j})$ are
cancelled
by those coming from the $x_k$ derivatives of ${\cal L}_n$ from the last term:
$({\pd V\over\pd\phi_k})({\pd{\cal L}_n\over\pd\phi_{ij}})\phi_{ijk}$.
In fact, we have explicitly:
$${\cal L}_{n+1} =-({\pd{\cal L}_n\over\pd\phi_{ij}}\phi_{ik}\phi_{jl}
   -{\cal L}_n\phi_{kl})M_{kl}\eqno(4.5)$$
where $M_{ij}={\pd~2V\over\pd\phi_i\pd\phi_j}$.

On the other hand, if $V$ is a homogeneous function of weight one depending
only upon $\phi_i$, and if ${\cal L}_n$ scales with a factor $F~\prime(\phi)$
under arbitrary field redefinitions $\phi\mapsto F(\phi)$, then
${\cal L}_{n+1}$ also scales with the same factor. Indeed,
under a functional replacement of $\phi$ by $F(\phi)$, the term
${\cal L}_n(\phi+\epsilon V)$ in (4.5) transforms as follows:
$$\eqalign{
{\cal L}_n(F(\phi)+\epsilon V(F(\phi)))=&{\cal L}_n(F(\phi)+\epsilon F'(\phi)V)
={\cal L}_n(F(\phi+\epsilon V)+\ \hbox{o}(\epsilon~2))\cr
=&F~\prime(\phi+\epsilon V){\cal L}_n(\phi+\epsilon V)+\
\hbox{o}(\epsilon~2).\cr}\eqno(4.6)$$
The first equality makes use of the homogeneity of $V$; the last of the scaling
property of ${\cal L}_n$.
Terms of order
$\epsilon~2$ may be safely omitted. Hence finally;
$${\cal L}_{n+1}(F(\phi))=F~\prime(\phi){\cal L}_{n+1}(\phi)+
{\cal L}_n(\phi)F~{\prime\prime}(\phi)(\phi_k{\pd V\over\pd\phi_k}-V).
\eqno(4.7)$$
However, the last term in (4.7) cancels, since the homogeneity of $V$ implies
implies the identities
$$\phi_k{\pd V\over\pd\phi_k}=V,\quad \sum_{j}M_{ij}\phi_j =0.\eqno(4.8)$$
Now consider the iterative sequence (4.4) with ${\cal L}_0(\phi_i)=V(\phi_i)$,
where $V$ is any  homogeneous function of weight one depending
only upon $\phi_i$.  Then all the members of the hierarchy scale with a factor
$F~\prime(\phi)$ and are recursively obtained from (4.5).

 The generic member of the hierarchy is then
$${\cal L}_n = VM_{i_1j_1}M_{i_2j_2}\cdots M_{i_nj_n}
\det\bigl|{\pd~2{\phi}\over\pd x_{i_p}\pd x_{j_q}}\bigr|,\eqno(4.9)$$
where a summation over all choices of $2n$ variables
$i_1\dots i_n,\ j_1\dots j_n$ out of $2d$ is implied, and the determinant is
that of the corresponding $n\times n$ matrix. The crucial point, as will be
explained in more detail in the second proof, is that in the generic case $M$
is a matrix of rank $d-1$ and the corresponding matrix of the cofactors of $M$
(the adjugate matrix, since $M$ is symmetric), is of rank 1.
 In consequence  (4.8) may be solved  to discover
$(\hbox{adj} M)_{ij}=\det(M)(M~{-1})_{ij}$  is
proportional to $\phi_i\phi_j$. Thus, for $n=d-1$ (4.9) may be written in the
dual manner
$${\cal L}_ {d-1}=  g(\phi_i)\sum_{i,j} \phi_i \bigl|{\pd~2{\phi}
\over\pd x_{i_p}\pd x_{j_q}}\bigr|~{-1}_{ij}\phi_j ,\eqno(4.10)$$
where $g(\phi_i)$ is a proportionality function which depends upon the details
of
${\cal L}_0$. The other factor is {\it independent of the starting Lagrangian}
and is the Universal Field Equation of the title, which is the generalisation
of the Bateman equation. In a similar manner ${\cal L}_d\equiv0$.

In the following paragraphs an alternative, more detailed proof is given which
depends upon the construction of a generating function for members of the
hierarchy.
\smallskip
\leftline{\it Theorem.}
\vskip 10pt
 Given any Lagrangian ${\cal  L}_0(\phi_i)$
which is a function of  $\phi_i$ alone, we have
$$\eqalign{{\cal L}(\phi_i,\phi_{ij};\lambda)=
\bigl[\exp\lambda {\cal L}_0{\cal E}]{\cal L}_0     &=(-1)~d{\cal L}_0(\phi_i)
\det\pmatrix{\lambda\phi_{ij}& {\gff}\cr{\gff}&-M_{kl}(\phi_i)\cr}\cr
                    &={\cal L}_0(\phi_i)\det[{\gff}+\lambda M\Phi].\cr}
\eqno(4.11)$$
Here $\Phi$ is the matrix with elements $\phi_{ij}$.
\vskip 10pt
 \leftline{\it Proof.}

The proof proceeds with the establishment of the following Lemma;
\vskip 10pt
\leftline{\it Lemma 2.}

$\det({\gff}+\lambda M\Phi)$ is a constant plus a divergence.
The easiest way to demonstrate this is to write the determinant as
$$\det({\gff}+\lambda M\Phi)=
(\delta_{i_1j_1}    +\lambda M_{i_1p_1}\Phi_{p_1j_1})
(\delta_{i_2j_2}+\lambda M_{i_2p_2}\Phi_{p_2 j_2})
\cdots\epsilon_{i_1i_2\cdots i_d }\epsilon_{j_1j_2\cdots j_d}.\eqno(4.12)$$
Recalling the definition of $M$, the right hand side can be expressed as
$$\eqalignno{
(\delta_{i_1j_1}+&\lambda\pd_{j_1}({\pd{\cal L}_0\over\pd\phi_{i_1}}))
(\delta_{i_2j_2}+\lambda M_{i_2p_2}\Phi_{p_2j_2})\cdots
\epsilon_{i_1i_2\cdots i_d}\epsilon_{j_1j_2\cdots j_d}&(4.13a)\cr
=& (\delta_{i_1j_1}+\lambda\pd_{j_1})\bigl(({\pd{\cal L}_0\over\pd\phi_{i_1}})
(\delta_{i_2j_2}+\lambda M_{i_2p_2}\phi_{p_2j_2})\cdots
\epsilon_{i_1i_2\cdots i_d}\epsilon_{j_1j_2\cdots j_d}\bigr).&(4.13b)\cr}$$
The expansion of (4.13b) yields (4.13a) together with a number of derivatives
which vanish on symmetry grounds. The non divergence part of (4.13b) is
$$(\delta_{i_2j_2}+\lambda M_{i_2p_2}\Phi_{p_2j_2})\cdots
\epsilon_{i_1i_2\cdots i_d}\epsilon_{i_1j_2\cdots j_d}.\eqno(4.14)$$
Proceeding iteratively, a further divergence can be extracted as in (4.13)
etc., until the only part which is not a divergence is a constant, $d!$.
Thus, as in lemma 1, the result follows that the term in
${\cal E}{\cal L}_0\det({\gff}+\lambda M\Phi)$ which is
proportional to ${\cal L}_0$ vanishes identically.

The terms  in ${\pd{\cal L}_0\over\pd\phi_i}$ in the calculation of
${\cal E}{\cal L}_0 \det({\gff}+\lambda M\Phi)$ are then
$$\eqalign{
({\pd{\cal L}_0\over\pd\phi_i})\pd_i\det({\gff}+\lambda M\Phi)
 &+({\pd_i{\cal L}_0}){\pd \over\pd\phi_i}\det({\gff}+\lambda M\Phi)\cr
 -[({\pd{\cal L}_0\over\pd\phi_k})\phi_{ijk}&
+2({\pd_i{\cal L}_0}) \pd_j]{\pd\det({\gff}+\lambda M\Phi)\over\pd\phi_{ij}}.
\cr}\eqno(4.15)$$
To evaluate the last term in the above expression two further lemmas are
required;
\vskip 10pt
 \leftline{\it Lemma 3.}
$$\eqalign{
\sum_{j,k}\phi_{jk}{\pd~2 \over\pd\phi_{ij}\phi_k}\det({\gff}+\lambda M\Phi)
&=\det({\gff}+\lambda M\Phi)\sum_{jkl}[{\gff}+\lambda M\Phi]~{-1}_{jl}
\lambda{\pd M_{li}\over\pd\phi_k}\Phi_{kj}\cr
&=\det({\gff}+\lambda M\Phi)\sum_{jkl}[{\gff}+\lambda M\Phi]~{-1}_{jl}
\lambda{\pd M_{lk}\over\pd\phi_i}\Phi_{kj}\cr
&={\pd \over\pd\phi_i}\det({\gff}+\lambda M\Phi).\cr}\eqno(4.16)$$
\vskip 10pt
 \leftline{\it Lemma 4.}
$$\eqalign{&{\pd~2 \over\pd\phi_{ij}\phi_{kl}}\det({\gff}+\lambda M\Phi)
  =\det({\gff}+\lambda M\Phi)\sum_{m,n}\lambda~2M_{mi}M_{nk}\cr
&\times \{[{\gff}+\lambda M\Phi)]~{-1}_{jm}
[{\gff}+\lambda M\Phi)]~{-1}_{ln}-[{\gff}+\lambda M\Phi)]~{-1}_{jn}
[{\gff}+\lambda M\Phi)]~{-1}_{lm}\}.\cr}\eqno(4.17)$$
Consequently
$$\sum_{j,k,l}\phi_{jkl}{\pd~2 \over\pd\phi_{ij}\phi_{kl}}
\det({\gff}+\lambda M\Phi)=0.\eqno(4.18)$$

Returning to expression (4.16), and using the chain rule\hfil\break
$\pd_i\mapsto \sum_j \phi_{ij}{\pd\over\pd\phi_j}+
\sum_{j,k}\phi_{ijk}{\pd\over\pd\phi_{jk}}$,
 and lemmas 3,4, the term in
$\phi_{ijk}$ cancels between the first and the third contribution to (4.15),
and also the remaining terms cancel, thus verifying that (4.15) vanishes.

Thus, finally the only non vanishing terms are those involving second
derivatives of ${\cal L}_0$ ,i.e.
${\pd~2{\cal L}_0\over\pd\phi_i\pd\phi_j}=M_{ij}$, giving
$$\eqalign{
&{\cal E}{\cal L}_0\det({\gff}+\lambda M\Phi)\cr
&={\pd~2{\cal L}_0
\over\pd\phi_k\pd\phi_l}\bigl(\phi_{ij}\delta_{ik}\delta_{jl}
-\phi_{ik}\phi_{jl}{\pd \over\pd\phi_{ij}}\bigr)\det({\gff}+\lambda M\Phi)\cr
 &= \hbox{tr}(M\Phi-[{\gff}+\lambda M\Phi]~{-1}\lambda M\Phi M\Phi)
 \det({\gff}+\lambda M\Phi)\cr
  &=\hbox{tr}([{\gff}+\lambda M\Phi]~{-1}M\Phi)\det({\gff}+\lambda M\Phi)\cr
  &={\pd\over\pd\lambda} \det({\gff}+\lambda M\Phi).\cr}\eqno(4.19) $$
This equation is a linear equation for $\det({\gff}+\lambda M\Phi)$
with solution
$$ \det({\gff}+\lambda M\Phi)=\exp(\lambda{\cal E}{\cal L}_0).\eqno(4.20)$$
where the right hand side stands for the result of the exponentiated operator
$\lambda{\cal EL}_0$ acting on 1.
This completes the proof of the theorem. What we have is then a generating
function for the sequence of equations of motion
${\cal L}_0{\cal E}{\cal L}_0\cdots{\cal E}{\cal L}_0$
arising from the iterated Lagrangians
${\cal L}_0{\cal E}{\cal L}_0{\cal E}{\cal L}_0\cdots{\cal E}{\cal L}_0$.

This sequence has a cohomological flavour to it, with ${\cal LE}$ playing the
role of an exterior derivative, and the sequence (4.2) being reminiscent of
de Rahm Cohomology. Also ${\cal E}\pd_j =0$ which implies that Lagrangians are
equivalent up to divergences.  In addition,
${\cal E}~2 =0$. One of the remarkable
features of this sequence is the fact that the highest derivatives which occur
are never more than second order. In establishing the theorem, no use has
been made of any homogeneity properties of ${\cal L}_0$. Nevertheless, the
iterative procedure terminates after the $d$-th stage, the determinant (4.20)
being a finite polynomial in $\lambda$ of degree $d$.

If we now suppose
${\cal L}_0$ to be homogeneous of weight $\alpha$, under the transformation
${\mit \Phi}=F(\phi)$ the generating function
${\cal L}(\phi_i,\phi_{ij};\lambda)$ behaves as
$$\eqalign{       {\cal L}({\mit \Phi_i},{\mit\Phi_{ij}};\lambda)=&
(F~\prime)~\alpha{\cal L}_0(\phi_i)det[\delta_{jk}
+\lambda(F~\prime)~{\alpha-1}
 \sum_{  m}M_{jm}(\phi_i)\phi_{mk}\cr  +&
\lambda(\alpha-1)(F~\prime)~{\alpha-2}F~{\prime\prime}{\pd{\cal L}_0
\over\pd\phi_j}\phi_k]\cr}.\eqno(4.21)$$
Hence, if $\alpha =1$ we have
$${\cal L}({\mit \Phi_i},{\mit\Phi_{ij}};\lambda)=
  F~\prime{\cal L}({\phi_i},{\phi_{ij}};\lambda)\eqno(4.23)$$
thus establishing the general covariance of all equations of the hierarchy
at a stroke.

The most important consequence of taking a homogeneous Lagrangian of
weight one  follows from the remark that $\sum_jM_{ij}\phi_j=0$, which
implies that $M$ is a matrix of at most rank $d-1$.

\vskip 10pt
\leftline{\it Theorem}
\vskip 10pt
 If ${\cal L}_0(\phi_i)$ is a homogeneous function of weight one, and the
matrix
$M_{ij}=  {\pd~2{\cal L}_0(\phi_k)\over\pd\phi_i\pd\phi_j} $ is of maximal rank
$d-1$ then the sequence of equations terminates after $d$ iterations. At the
penultimate step there is an universal equation
$\sum_{i,j}\phi_i(\hbox{adj}\Phi)_{ij}\phi_j=0$, where $\hbox{adj}\Phi$
denotes the adjugate matrix to $\Phi$. This equation may be regarded as the
generalisation of the Bateman equation to $d>2$ dimensions.
\vskip 10pt
\leftline{\it Proof}
\vskip 10pt
 The term in $\lambda~d$ in the expansion of
${\cal L}(\phi_i,\phi_{ij};\lambda)$ is $(-1)~d{\cal L}_0\det M\det \Phi$
which vanishes as $M$ is of rank $d-1$. The coefficient of $\lambda~{d-1}$
in   $\cal L$ is $(-1)~d{\cal L}_0\hbox{tr}(\hbox{adj}M\hbox{Adj}\Phi).$
The universality of this expression follows from the observation that
$(\hbox{Adj}M)_{ij}=g(\phi_k)\phi_i\phi_j$. If the rank of $M$ is less than
$d-1$, the expansion terminates at an earlier stage.This completes the proof.

As a final remark, using the results of section 3 we shall show that
under the conditions of this theorem the class of functions implicitly
defined by
$$ \sum~d_{j=1}x_jF_j(\phi) =c,\eqno(4.24)$$
constitute a subset of solutions to each member of the hierarchy of equations.
A simple calculation shows that
$$\sum_kM_{ik}(\phi_i)\Phi_{kj}={-1\over S}\sum_kM_{ik}(F_i)F~\prime_kF_j,
\eqno(4.25)$$
so that $\hbox{tr}(M(\phi_i)\Phi)~k=0$ and

$${\cal L}(\phi_i,\phi_{ij};\lambda)={\cal L}_0\exp\sum_{k=1}~{\infty}
(-1)~{k-1}{\lambda~k\over k}\hbox{tr}(M\Phi)~k={\cal L}_0.\eqno(4.26)$$
This establishes the class of solutions for all members of the hierarchy.
Note also that for the universal equation, any function
$\phi(x_1,\dots\hat{x}_k,\dots x_d)$ which depends on all variables except
$x_k$ is automatically a solution.

Except for $d=2$, (4.24) does not define the general solution to the
universal equation (4.3). Any homogeneous function $\phi(x_i)$ of zero degree
is always a solution, for example. Also if $\phi~{(d)}(x_i)$ is a solution
to the $d$-dimensional equation, $\phi~{(d)}(x_i+h_i(x_{d+1}))$ is a solution
to the $d+1$-dimensional equation, with $h_i(x_{d+1})$ being arbitrary
functions of $x_{d+1}$. Finally, the following products,
$$ \phi(x_i)= \prod~N_{n=1}\theta_n~{\alpha_n},\quad \theta_n=\sum_ia_{ni}x_i
+b_n,\eqno(4.27)$$
define solutions for any parameters $\alpha_n,\ a_{ni},\ b_n$ when $N<d$.
For $N=d$, we require additionally either $\sum_n \alpha_n=0$ or det$a_{ni}=0$.
\vskip 10pt
\leftline{\bf 5. Generalisation to Multiple Fields.}
\vskip 10pt
The extension of the results of the last section to the case ${\cal D}>1$ is
not
straightforward. First of all, we shall consider the case where
$d={\cal D}+1$ and demonstrate the existence of a univesal equation, which
generalises the Bateman equation. Introduce the notation
$J_{i,j,\dots,k} =\det{\pd\phi~a\over\pd x_i}$ for the jacobian of
$\phi~a,\phi~b,\dots,\phi~l$ with respect to $x_i,x_j,\dots,x_k$. In general,
for ${\cal D}<d$ there are ${d \choose{\cal D}}$ such jacobians, but in the
case
where $d={\cal D}+1$ there are precisely $d$ of them, which may be labelled
by the index of the variable which is ommitted, viz. $J_i$.
Now suppose that ${\cal L}_0$ is considered as a function of the jacobians
$J_i$.

Because,  under
a field redefinition of the form of equations (3.12), $J_i$ transforms into
$\det A J_i$, the restriction of ${\cal L}_0$  to the class of  homogeneous
functions of weight one is similar to that for a single scalar field.
The equations of motion are simply
$$ \eqalign{
{\cal E}~a{\cal L}_0 =&{\pd\over\pd x_i}{\pd{\cal L}_0\over\pd\phi~a_i}\cr=&
({\pd{\cal L}_0\over\pd J_j}
{\pd~2J_j\over\pd\phi~a_i\pd\phi~b _k} \phi~b _{ik}
+{\pd~2{\cal L}_0\over\pd J_j\pd J_l}{\pd J_j\over\pd\phi~a_i}
{\pd J_l\over\pd\phi~b _k}\phi~b_{ik})
=0.\cr}
\eqno(5.1)$$
The first term vanishes from symmetry considerations. From the homogeneity of
${\cal L}_0 $ we have
$$\sum~d_k {\pd~2{\cal L}_0\over\pd J_j\pd J_k}J_k = 0.\eqno(5.2)$$
But since $\sum_k J_k\phi~a_k=0,\ \forall a$, this together with symmetry,
implies that
$${\pd~2{\cal L}_0\over\pd J_i\pd J_j}= \sum_{a,b}\phi~a_i d~{a,b}\phi~b_j,
\eqno(5.3)$$
for some functions $d~{a,b}$. Inserting this representation into (5.1) and
using \hfil\break
$\sum_j{\pd J_j\over\pd\phi~a_i}\phi~c_j=-\delta~{a,c}J_i$
the equations of motion become, assuming $d~{a,b}$ is invertible, as in the
generic case:
$$\sum_{i,k}J_iJ_k\phi~a _{ik}=0.\eqno(5.4)$$
This is a direct generalisation of the Bateman equation to more than two
dimensions, and is the universal set of field equations when $d={\cal D}+1$,
being independent of the starting Lagrangian.

In the more general case $d={\cal D}+n$, there is no longer a one to one
correspondence between jacobians and derivatives. Suppose the jacobians are
labelled by the indices corresponding to the indices of the $n$ ommitted
derivatives; $J_{i_1,\dots,i_n}$. We are led to conjecture the following set
of equations as universal (all lower indices summed):
$$J_{i_1,\dots,i_n}J_{j_1,\dots,j_n}\det
\pmatrix{\phi~{a_1}_{i_1,j_1}&\ldots&\phi~{a_1}_{i_1,j_n}\cr
         \vdots          &\ddots&\vdots \cr
         \phi~{a_n}_{i_n,j_1}&\ldots&\phi~{a_n}_{i_n,j_n}\cr}=0.\eqno(5.5)$$
Note that the indices $a_j$  which enumerate the fields $\phi$ in the above
equations are free. Since they occur symmetrically, the number of such
equations is $d-1\choose{\cal D}-1$. Thus they form an overdetermined set.
However, under a field redefinition of the form
$ {\mit\phi}~a   =F~a(\phi~b)$ it is straightforward to verify that
all equations (5.5) transform into linear combinations of each other, and thus
form an invariant set.
We have not yet succeeded in deriving (5.5) from a hierarchy principle. An
alternative expression for (5.5) is as follows:
$$\sum(\prod_{b_k}\phi~{b_k}_{i_l})(\prod_{c_k}\phi~{c_k}_{j_l})
(\prod_{a_k}\phi~{a_k}_{i_mj_m})\epsilon_{i_1\dots i_d}\epsilon_{i_1\dots i_d}
=0.\eqno(5.6)$$
It is also straightforward to verify that the implicit solution defined by
$$\sum_i   x_iF~a_i (\phi~b) = c~a,\quad F~a_j\ \hbox{ arbitrary}.
\eqno(5.7)$$
is a solution to (5.6). As in section 3, (5.7) implies a representation
for $\phi~a_{ij}$ of the form
$$\phi~a_{ij} = \sum_b(S~a_{i,b}\phi~b_j+S~a_{j,b}\phi~b_i)\eqno(5.8)$$
Inserting this expression in (5,6) every term in the expansion is seen to
contain at least one pair of derivatives of the type $\phi~a_i\phi~a_j$.
Such terms vanish by antisymmetry of $\epsilon$. Thus the solution space is
non-empty.

\vskip 10pt
\leftline{\bf Conclusions.}
\vskip 10pt
The main result of this paper is the discovery of a hierarchical
system of iterated Lagrangians and associated equations of motion which
terminate in a universal equation, independent of the initial Lagrangian.
This universal equation posesses several remarkable properties. It is invariant
under $GL(d)$ transformations of the base space, though the starting Lagrangian
need have no such symmetry. It possesses an infinite number of conservation
laws in virtue of the fact that at the previous level in the hierarchy, there
are an infinite number of Lagrangians whose variation yields it. This gives the
possibility that the equation may be completely integrable. As it stands
some classes of solution were recorded in section 4, but the general solution
remains to be discovered.

The extension of this idea to a higher dimensional target space, in section 5
is as yet only partially realised, where a universal equation is proved for
the situation where the base space is just one dimension greater than the
target  space, as for the original Bateman equation. A set of
universal equations invariant under field redefinitions
is conjectured in the general case. This set  is somewhat novel in being
overdetermined. Nevertheless the solution space is non empty. Work remains
to be done in finding the correct hierarchy principle which leads to such
equations.

This investigation suggests also the possible existence of an extension to the
membrane situation, where the target space has greater dimension than the
base space. In the usual membrane equations, the second derivatives of the
field quantities (target space co-ordinates) appear once and once only.
Our results suggest the existence of similar universal equations  which are
multinomials in the second derivatives, the invariance under field
redefinitions being replaced by reparametrisation invariance. Work is in
progress in the hope to realise thereby a fully integrable membrane equation.

One aspect of the Bateman equation which has not been successfully generalised
is the property (B) of section 2, in which it appeared as the restriction of
solutions to the free wave equation in 2+1 dimensions to the subspace upon
which the gradient vector is null. Further work in this direction might prove
illuminating as it corresponds to a theory   in which the Lagrangian itself is
constrained to be zero, reminiscent of the situation with topological
Lagrangians.

 It may be that the iterative principle of constructing sequences of
Lagrangians and equations of motion initiated here might be capable of many
further generalisations;  to include fields themselves, for example.
The geometrical significance of this procedure also begs for elucidation;
the gauge nature of the theories so constructed  requires understanding.
\vskip 10pt
\leftline{\bf Acknowledgement}
\vskip 5pt
This research was supported by the S.E.R.C. award of a Research
Assistantship to J.Govaerts and a Visiting Fellowship to A. Morozov.
\vfill\eject
\centerline{\bf REFERENCES}
\frenchspacing
\item{[1]}Witten E.,{\it Comm. Math. Phys.}\ {\bf  117}\ (1988)\ 353;
ibid.\ {\bf 118}\ (1988)\ 411.
\item{[2]}Nambu  Y.,{\it Lectures At Copenhagen Symposium} (1970).
\item{[3]}Goto   T.,{\it Prog. Theor. Physics} {\bf 46}\ (1971)\ 1560.
\item{[4]}Bateman H., {\it Proc. Roy.Soc.London}\  {\bf A 125}\ (1929)\
598-618.
\item{[5]}Garabedian P.R., {Partial Differential Equations}\ Wiley
(1964) p.517.
\item{[6]}Olver  E.,{\it Applications of Lie Groups to Differential Equations}
\ Graduate Texts in Mathematics, {\bf 107}, Springer Verlag (1986), p.252.

\end